\documentclass[preprint,aps,prd,longbibliography,tightenlines,toc=flat]{revtex4-2}

\usepackage{graphicx} 
\usepackage{dcolumn}  
\usepackage{colordvi}
\usepackage{color}
\usepackage{epstopdf}
\usepackage{amsmath}
\usepackage{amssymb}
\usepackage{url}
\graphicspath{{ps}}
\usepackage{hyperref}

\usepackage{amsmath}
\usepackage{makecell}

\usepackage[dvipsnames]{xcolor}
\usepackage{float}
\usepackage{subcaption}
\usepackage{tabularx, ragged2e} 
\usepackage{adjustbox}
\usepackage{booktabs}
\usepackage{caption}
\usepackage{siunitx}
\usepackage{multirow}
\usepackage{listings}

\usepackage{cleveref}
\creflabelformat{equation}{#2\textup{#1}#3}

\definecolor{codegreen}{rgb}{0,0.6,0}
\definecolor{codegray}{rgb}{0.5,0.5,0.5}
\definecolor{codepurple}{rgb}{0.58,0,0.82}
\definecolor{backcolour}{rgb}{0.95,0.95,0.8}
\lstdefinestyle{mystyle}{
    backgroundcolor=\color{backcolour},   
    commentstyle=\color{codegreen},
    keywordstyle=\color{magenta},
    numberstyle=\tiny\color{codegray},
    stringstyle=\color{codepurple},
    basicstyle=\ttfamily\scriptsize,
    breakatwhitespace=false,         
    breaklines=true,                 
    captionpos=b,                    
    keepspaces=true,                 
    numbers=none,
    numbersep=5pt,                  
    showspaces=false,                
    showstringspaces=false,
    showtabs=false,                  
    tabsize=2,
}
\usepackage{adjustbox}

\usepackage{titlesec}
\titlespacing*{\subsection}{0pt}{1.1\baselineskip}{\baselineskip}
\titlespacing*{\section}{0pt}{1.1\baselineskip}{\baselineskip}
\titleformat*{\section}{\normalsize\bfseries}
\titleformat*{\subsection}{\small\bfseries}

\renewcommand{\thesection}{\arabic{section}}
\renewcommand{\thesubsection}{\thesection.\arabic{subsection}}

\makeatletter

\renewcommand{\p@subsubsection}{\thesubsection.}
\renewcommand{\p@subsection}{}
\renewcommand{\@seccntformat}[1]{%
  \csname the#1\endcsname
  \ifnum\pdfstrcmp{#1}{subsubsection}=0 .\fi
  \quad}
\makeatother

\input belle2sym.tex

\begin{document}

\def\belletwo {\it {Belle II}}

\vspace*{-3\baselineskip}
\resizebox{!}{3cm}{\includegraphics{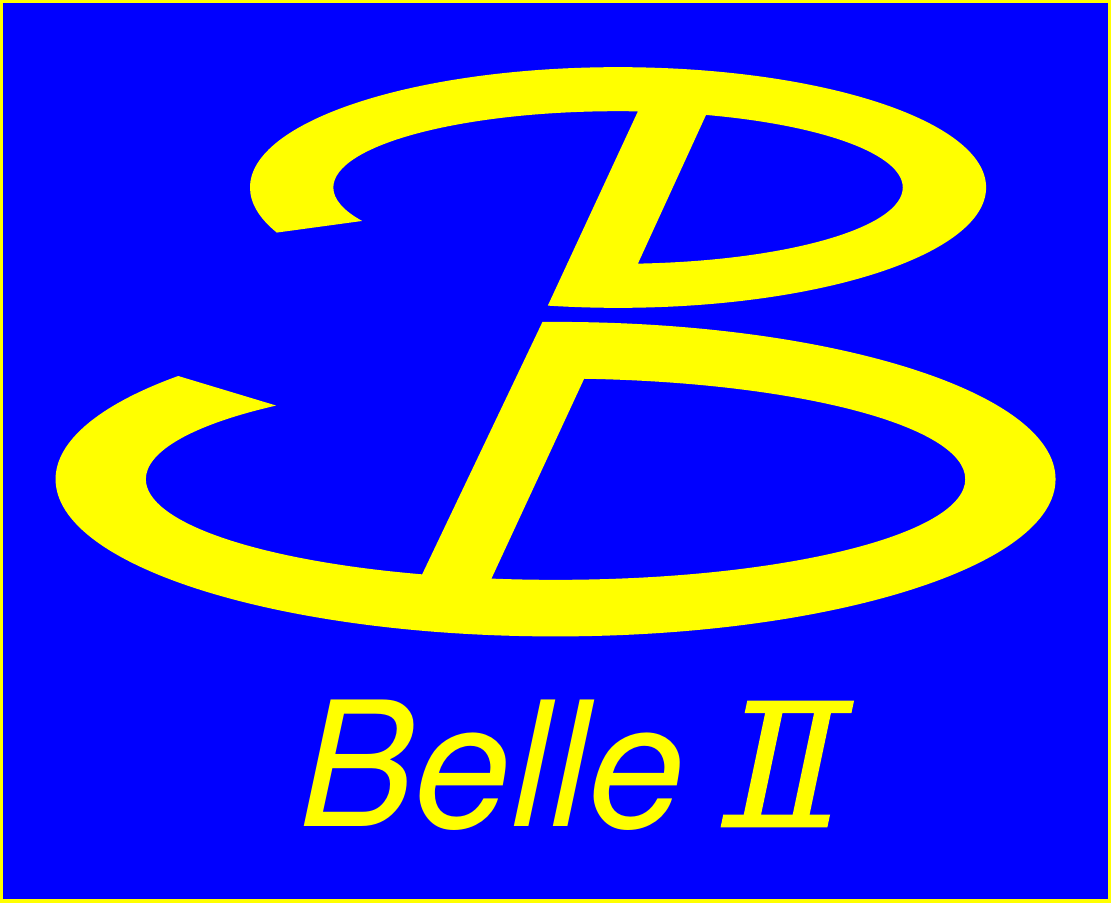}}

\vspace*{-5\baselineskip}
\begin{flushright}
BELLE2-CONF-PH-2022-018\\
\today
\end{flushright}

\title {\quad\\[0.5cm] Measurement of the photon-energy spectrum in inclusive \BtoXsgamma decays identified using hadronic decays of the recoil $B$ meson in 2019–2021 Belle II data}

\collaboration{The Belle II Collaboration}
  \author{F. Abudin{\'e}n}
  \author{I. Adachi}
  \author{K. Adamczyk}
  \author{L. Aggarwal}
  \author{P. Ahlburg}
  \author{H. Ahmed}
  \author{J. K. Ahn}
  \author{H. Aihara}
  \author{N. Akopov}
  \author{A. Aloisio}
% ...


\begin{abstract}

We measure the photon-energy spectrum in radiative bottom-meson ($B$) decays into inclusive final states involving a strange hadron and a photon. We use SuperKEKB electron-positron collisions corresponding to $189~\invfb$ of integrated luminosity collected at the \FourS resonance by the Belle II experiment. The partner \B candidates are fully reconstructed using a large number of hadronic channels. The \BtoXsgamma partial branching fractions are measured as a function of photon energy in the signal $B$ meson rest frame in eight bins above 1.8 GeV. The background-subtracted signal yield for this photon energy region is $343 \pm 122$ events. Integrated branching fractions for three photon energy thresholds of $1.8~\gev$, $2.0~\gev$, and $2.1~\gev$ are also reported, and found to be in agreement with world averages.

\end{abstract}


\maketitle

{\renewcommand{\thefootnote}{\fnsymbol{footnote}}}
\setcounter{footnote}{0}

\clearpage
\section{Introduction}

Flavour changing neutral currents (FCNCs) are only allowed in the Standard Model (SM) via loop processes and are therefore highly suppressed \cite{Misiak:2020vlo}. The \BtoXsgamma FCNC decays occur via radiative $b\rightarrow s$ transitions, where $B$ denotes charged and neutral bottom-mesons, and $X_s$ denotes all available final states containing net strangeness. These processes are particularly sensitive to non-SM effects \cite{Misiak:2017bgg}.
In addition, their photon-energy spectrum offers access to various interesting parameters, such as the mass of the $b$ quark and the function describing its motion inside the $B$ meson \cite{RevModPhys.88.035008, simba}.

We present an inclusive measurement using \BtoXsgamma decays identified in $\Upsilon(4S) \to B\overline{B}$ events in which the partner $B$ meson is reconstructed in its hadronic decays (hadronic tagging). This approach is complementary to the untagged or lepton-tagged (see e.g., \cite{BaBar:2012fqh}) and sum-of-exclusive (e.g., \cite{Belle:2014nmp}) methods because it has different sources of systematic uncertainty. In addition, tagging provides a purer sample and the kinematic information from the partner-$B$ meson gives direct access to observables in the signal-$B$ meson rest frame. We denote the photon energy in the signal-$B$ meson rest frame as \EB.  In this paper, the minimum \EB photon energy threshold is 1.8 GeV. The inclusive analysis does not distinguish between contributions from $b\ra\d\g$ and $b\ra\s\g$ processes, therefore the much smaller $b\ra\d\g$ contribution is subtracted from the final results with a shape determined from simulation.






\section{Belle II detector}

The Belle II \cite{Belle-II:2010dht} detector is designed to reconstruct the final states of electron-positron collisions at center-of-mass energies at or near the \FourS meson mass. The colliding \epem beams are provided by the SuperKEKB collider \cite{AKAI2018188} at KEK in Tsukuba, Japan. The detector has collected physics data since 2019. Belle II consists of several detector subsystems arranged cylindrically around the beam pipe. In the Belle II coordinate system, the $x$ axis is defined to be horizontal and points to the outside of the tunnel for the accelerator's
main rings, the $y$ axis is vertically upward, and the $z$ axis is defined in the direction of the electron beam. The azimuthal angle, $\phi$, and the polar angle, $\theta$, are defined with respect to the $z$ axis. Three regions in the detector are defined based on $\theta$: forward endcap (\mbox{$12^{\circ}<\theta<31^{\circ}$}), barrel (\mbox{$32^{\circ}<\theta<129^{\circ}$}) and backward endcap (\mbox{$131^{\circ}<\theta<155^{\circ}$}).

The Belle II vertex detector is designed to precisely determine particle decay vertices. It is the innermost subsystem, and consists of a silicon pixel detector and a silicon strip detector. Surrounding the vertexing subsystems is the central drift chamber,
which is used to measure charged-particle trajectories (tracks) to determine their charge and momentum. It also provides important particle-identification information by measuring the specific ionisation of charged tracks. Further particle identification is provided by the time-of-propagation detector and the aerogel ring-imaging Cherenkov detector, which cover, respectively, the barrel and the forward endcap regions of Belle II. Photons and electrons are stopped and their energy deposits (clusters) are read out by the CsI(Tl)-crystal electromagnetic calorimeter. The photon-energy resolution of the ECL is better than 20~\mev for photons above 1~\gev.
All the inner components are surrounded by a superconducting solenoid, which provides a uniform axial 1.5 T magnetic field. The $K^0_L$ and muon detector, composed of plastic scintillators and resistive-plate chambers, is the outermost subsystem of Belle~II.

\section{Data sets}\label{sec:datasets}

The results presented here use a data sample corresponding to 189~\invfb of integrated luminosity collected at an energy corresponding to the \FourS mass. In addition, an off-resonance data set corresponding to 18~\invfb collected 60~\mev below the \FourS resonance is used to validate the \epem\ra\qqbar simulation. Here $q$ is used to indicate $u,~d,~s$ and $c$ quarks.

The relevant background and signal processes are modeled using large samples simulated through the Monte Carlo~(MC) method corresponding to \mbox{1.6 \invab} of \epem\ra\qqbar events (generated by KKMC~\cite{Ward:2002qq}, interfaced to PYTHIA~\cite{Sjostrand:2007gs}) and  \mbox{\FourS\ra\BzBzb, \BpBm} events (generated by EVTGEN~\cite{Ryd:2005zz}). The detector response is simulated using Geant~4~\cite{Agostinelli:2002hh}.

In addition, inclusive \BtoXsgamma signal distributions are generated using \texttt{BTOXSGAMMA}, the EVTGEN implementation of the Kagan-Neubert model \cite{Kagan:1998ym}, with values of the model parameters taken from Ref. \cite{simba}. The inclusive model, by construction, does not reproduce the resonant structure of the $\b\ra\s\g$ transitions. Therefore, the $\B\ra K^*(892)\gamma$ sample (later denoted as $\B\ra K^*\g$) generated by EVTGEN is also used, as it dominates the higher end of the \Egamma spectrum. The \BtoXsgamma and $\B\ra\Kstar\gamma$  signal simulations are combined using a ``hybrid-model'', inspired by Ref. \cite{hybrid_model}. The full spectrum is modelled by the combination of the two simulated signal samples. A set of hybrid \EB intervals (bins) is defined and the \BtoXsgamma spectrum is scaled in each bin to match the partial branching fraction of the combined \BtoXsgamma and $\B\ra\Kstar\g$ decays with the expected value. The hybrid signal model is used in the selection optimisation, efficiency determination, and unfolding procedure.


All the data sets are analysed using the Belle II analysis software framework \cite{Kuhr:2018lps}.

\section{Analysis overview}\label{sec:analysis_flow}

A sample of tagged $B$ mesons is first reconstructed in their hadronic decays, and a high-energy photon from the other $B$ meson is selected. Details of the tagging algorithm are described in \Cref{sec:tag_side_reconstruction}. The selection procedures to suppress photon candidates from background processes are given in \Cref{sec:signal_side_reconstruction,sec:qqbar_selection}. They are optimised in simulation, simultaneously, by maximising the figure of merit of Ref.~\cite{Punzi}. The final best tag-candidate selection is summarised in \Cref{sec:best_tag_candidate}.
The fitting of sample composition is described in \Cref{sec:fitting_procedure}. The procedure to remove the remaining background contamination after the fit is given in \Cref{sec:post_fit}. The reconstructed \BtoXsgamma event yields in bins of \EB are unfolded (\Cref{sec:unfolding}). The corresponding uncertainties are discussed in \Cref{sec:uncertainties}. The final results of the analysis are presented in \Cref{sec:results}.

The analysis is fully optimised on simulation. Control regions are used to check the validity of the background suppression before examining the signal region.

\section{Event reconstruction and selection}\label{sec:selection}


\subsection{Tag side reconstruction}\label{sec:tag_side_reconstruction}

In each event, one $B$ meson candidate is fully reconstructed and used as a tag for the recoiling signal $B$ meson candidate. The tag-side \B meson is reconstructed using the full-event-interpretation algorithm (FEI) \cite{FEI}, which reconstructs hadronic $B$ decays from thousands of subdecay chains. The algorithm starts by combining track and ECL cluster information to reconstruct final-state candidate particles, such as electrons, muons, photons, charged pions, and kaons. In the next step, those are combined to form intermediate particles
such as \piz, $K_S^0$, $D^{(*)}$, and \jpsi candidates. The intermediate or final-state particles are combined to form \B candidates in 36 \Bp and 32 \Bz hadronic modes. For each reconstructed \B-meson candidate, the algorithm outputs a probability-like score, \feiProb. Correctly reconstructed $B$-meson candidates have a score close to one, whereas non-$B$ and misreconstructed candidates tend to have a score close to zero.

For tag-candidate reconstruction, track-quality requirements are imposed \cite{BERTACCHI2021107610}. The longitudinal distance of closest approach of each track from the detector center is required to be $|z_0|<2.0~\cm$. A similar criterion for the distance in the transverse plane, $|d_0|<0.5~\cm$ is also applied. Only charged particles with transverse momenta, $p_T$, higher than 0.1~\gevc are selected. Furthermore, an event must have at least three tracks passing these selections. Similarly, three or more isolated clusters with $17<\theta<150^{\circ}$ and $E>0.1~\gev$ are required in the ECL for each event. The total energy deposited in the ECL should be within $2$ and $7~\gev$. Only events with at least $4~\gev$ of measured energy in the Belle II detector are retained. The tag-side \B candidates are required to have $\feiProb>0.001$ and beam-constrained mass, $\Mbc>5.245~\gevcc$, defined as

\begin{equation}\label{eq:mbc}
\Mbc = \sqrt{(\sqrt{s}/2)^2 - p_{\mathrm{tag}}^2},
\end{equation}

\noindent where $\sqrt{s}$ is the collision energy and $p_{\mathrm{tag}}$ is the reconstructed momentum of the tag-candidate in the CM frame. Furthermore, a $|\Delta E|<0.2~\gev$ requirement is imposed on the energy difference, defined as 

\begin{equation}
    \Delta E = E_{\mathrm{tag}} - \sqrt{s}/2,
\end{equation}

\noindent where $E_{\mathrm{tag}}$ is the energy of the tag-candidate reconstructed in the CM frame.

\subsection{Signal-side selection}\label{sec:signal_side_reconstruction}

Using the kinematic properties of the reconstructed tag-side meson and the beam-energy constraint, the photon-candidate energy is inferred in the signal-\B meson rest-frame. All candidates in the range of $\EB>1.4~\gev$ are considered. The highest-energy photon in each event is taken as the signal-photon candidate. Studies on simulated signal events show that this selects the correct \BtoXsgamma candidate in more than 99\% of cases. Cluster-timing information, derived from a waveform fit to the signal collected in the most energetic crystal of the cluster, is associated to each photon. In order to suppress the large background from out-of-time beam-background clusters or photons associated with low-quality waveform fits, the cluster timing, measured with respect to the collision time, is used. It is required neither to exceed $200~\ns$ nor twice the cluster-time uncertainty. This selection introduces about a 2\% signal efficiency loss while reducing beam background to negligible levels.

The resulting sample is dominated by photon candidates originating from asymmetric $\piz\ra\g\g$ and $\eta\ra\g\g$ decays. This background is suppressed by vetoing $\pi^0$ and $\eta$ decays. Signal-photon candidates are kinematically combined with lower-energy photons in the event and a quantitative measure of compatibility with $\piz\ra\g\g$ or $\eta\ra\g\g$ decays is associated to the combinations. The compatibility is determined using a multivariate statistical-learning algorithm that uses the diphoton invariant mass, helicity, and properties of the less-energetic photon, such as energy, polar angle, and smallest ECL cluster-to-track distance. Another statistical-learning algoritm uses Zernike moments to quantify the ECL-photon cluster shape to disentangle misidentified photons from signal clusters. The background suppression of these selections is investigated using the off-resonance data. Furthermore, samples with inverted \piz and $\eta$ suppression requirements are used to form large samples of background-like events.

\subsection{Suppression of \qqbar events}\label{sec:qqbar_selection}

Photon candidates from background \epem\ra\qqbar events make up most of the selected sample. A dedicated boosted-decision-tree classifier is trained to suppress these events. The training is performed on randomly selected sets of $10^5$ simulated \qqbar events and $10^5$ simulated \BtoXsgamma events that pass the requirements described in \Cref{sec:tag_side_reconstruction,sec:signal_side_reconstruction}. The input features for the classifier are tag-side \B kinematic parameters, such as modified Fox-Wolfram moments~\cite{KSFW_moments}, \B decay vertex parameters, CLEO cones \cite{CLEO_cones}, and thrust \cite{BaBar:2014omp}. For each variable, we require minimal correlations with \EB and \Mbc in order not to bias the inclusive spectrum. Furthermore, each variable distribution in off-resonance data is compared to \epem\ra\qqbar simulation. Only those showing good data-simulation agreement are used for the training. The classifier outputs a probability score, $\mathcal{P}_{\mathrm{BDT}}$, for each event to be classified as \epem~\ra~\qqbar or \epem~\ra~\BB where one of the \B mesons decays as \BtoXsgamma. 

\subsection{Best tag-side candidate selection}\label{sec:best_tag_candidate}

After all the selections described in \Cref{sec:selection}, about 50\% of events have a unique tag-side candidate, based on the hybrid signal model. In 25\% of cases there are two tag-side candidates remaining, and in 10\% of cases -- three. The probability to have more than one tag-side candidate decreases rapidly and is lower than 1\% for 5 or more candidates. If more than one tag-side candidate is present in an event, only the one with the highest \feiProb is retained. 

\subsection{Selection efficiency}

The signal efficiency is decomposed as a product of the FEI tagging efficiency and the signal-selection efficiency. Signal-selection and tagging efficiencies are calculated using the simulated hybrid model. The tagging efficiencies are calculated by comparing $B$ yields in simulation before and after the FEI algorithm is applied. After applying the FEI calibration factors of $\mathcal{C}_{B^+}=0.66 \pm 0.02$ and $\mathcal{C}_{B^0}=0.67 \pm 0.02$ for charged and neutral modes, respectively, a tagging efficiency of $(0.44\pm0.02)\%$ is found. The calibration factors account for differences in the tag-side reconstruction efficiency between data and simulation and are derived in an independent study of hadronic-tagged $B\ra X\ell\nu$ decays \cite{Belle-II:2020fst}. The signal-selection efficiency is calculated as the fraction of tagged-signal yield that meet the requirements of \Cref{sec:signal_side_reconstruction}. It increases approximately linearly from 45\% to 63\% with $\EB~\in~[1.8,2.6]~\gev$, with a total uncertainty of about 10\% in each bin. Background events from $\epem\ra\qqbar$ are suppressed by 99.5\%, while the generic $\epem\ra\BB$ background is reduced by 93\%.

\section{Fit of sample composition}\label{sec:fitting_procedure}

The tag-side \Mbc distribution is fit to determine the yields of $\B$ mesons that provide good kinematic constraints on the signal side, the remaining \qqbar events, and combinatorial \BB events. The sample is divided in 11 bins of \EB: three 200-$\si{\MeV}$-wide bins for the 1.4--2.0~\gev range; seven 100-$\si{\MeV}$-wide bins for the 2.0--2.7~\gev region; and a single $\EB>2.7~\gev$ bin. The first two bins and the last one are chosen as control regions for the fit due to expected large background or low signal yield. The signal region is therefore defined as \mbox{$1.8<\EB<2.7~\gev$}.

The model for the fit of sample composition is determined using simulation. The simulated sample is split into three components: correctly reconstructed (peaking) \BB events, \qqbar decays, and combinatorial background from \BB decays. `Peaking' henceforth is used generically to denote the resonant behaviour in \Mbc of correctly reconstructed tag-side $B$ decays. These components have distinct shapes in \Mbc, which are parameterised using probability distribution functions (PDFs). To extract the yield of peaking \BB tags, a Crystal Ball function is used~\cite{CrystalBall:1986xj}. This function is the sum of a peaking Gaussian part and a polynomial tail. The \qqbar decays are described by an ARGUS function~\cite{ALBRECHT1990278}. Combinatorial \BB background is described by a fifth-order Chebyshev polynomial. Studies on simulation show that a lower-order polynomial is insufficient to accurately describe the \Mbc shape of this component. 



\begin{figure}[htbp!]
    \centering
    
    \includegraphics[width=0.45\textwidth]{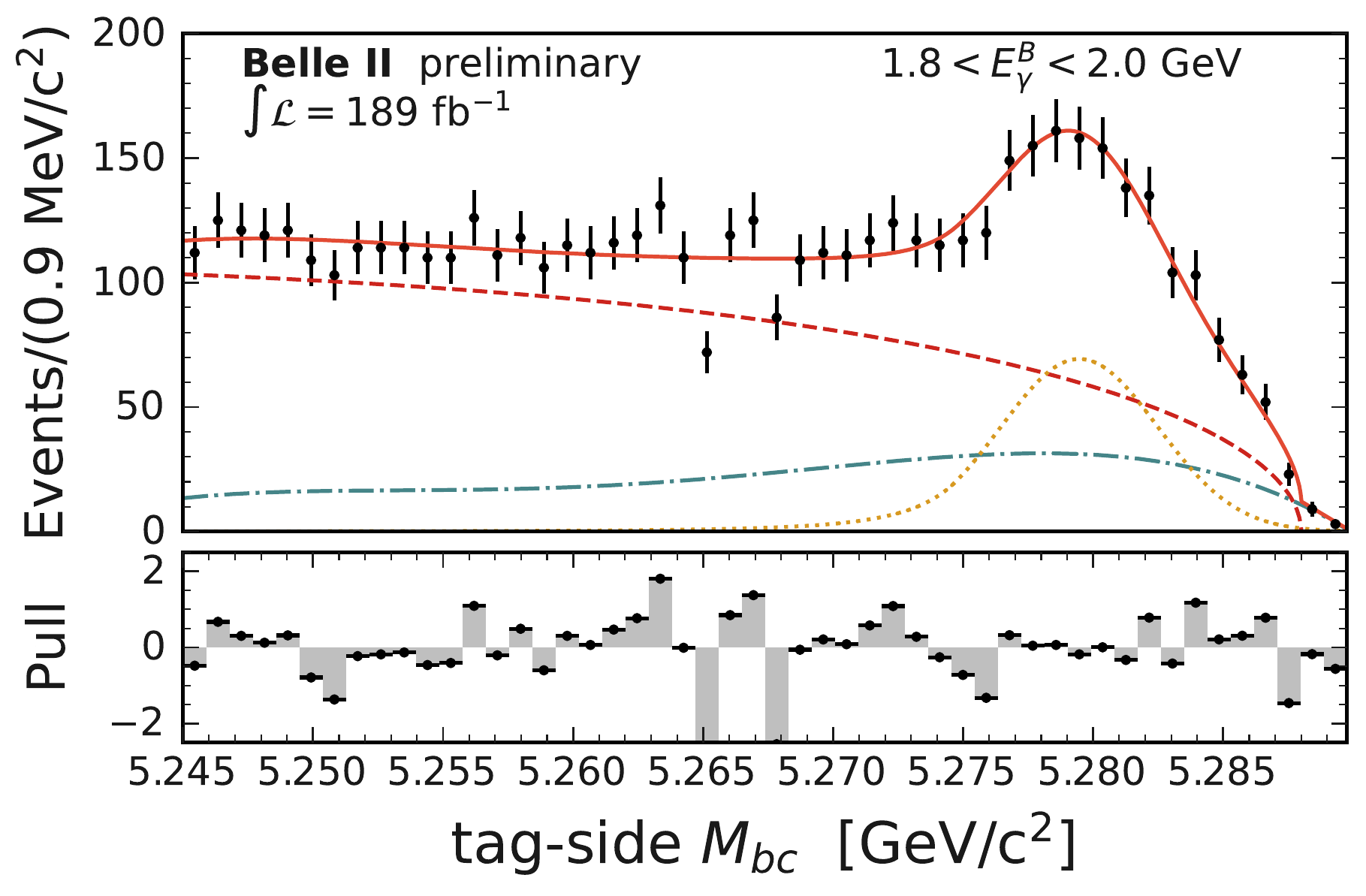}
    \includegraphics[width=0.45\textwidth]{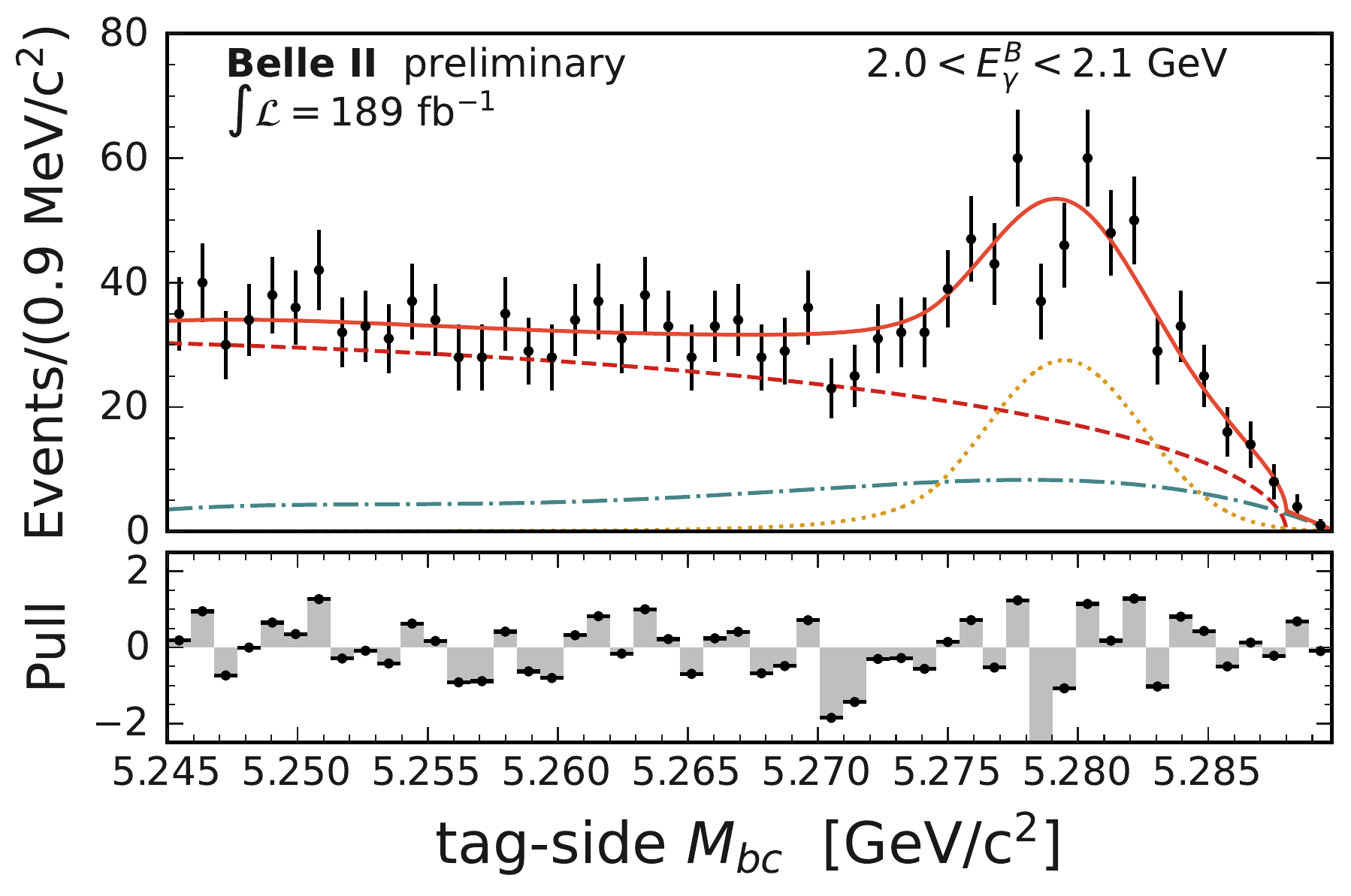}
    \includegraphics[width=0.45\textwidth]{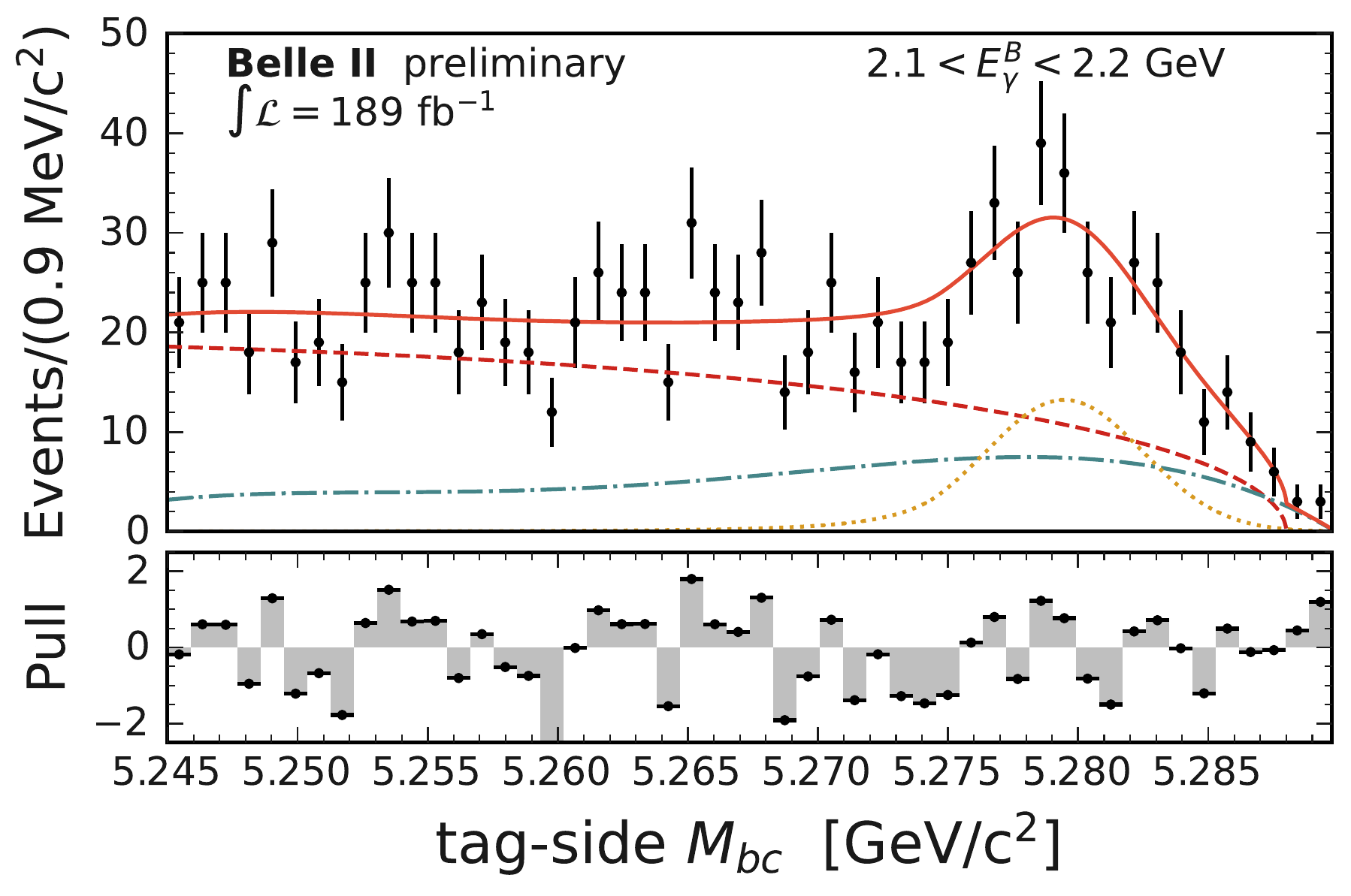}
    \includegraphics[width=0.45\textwidth]{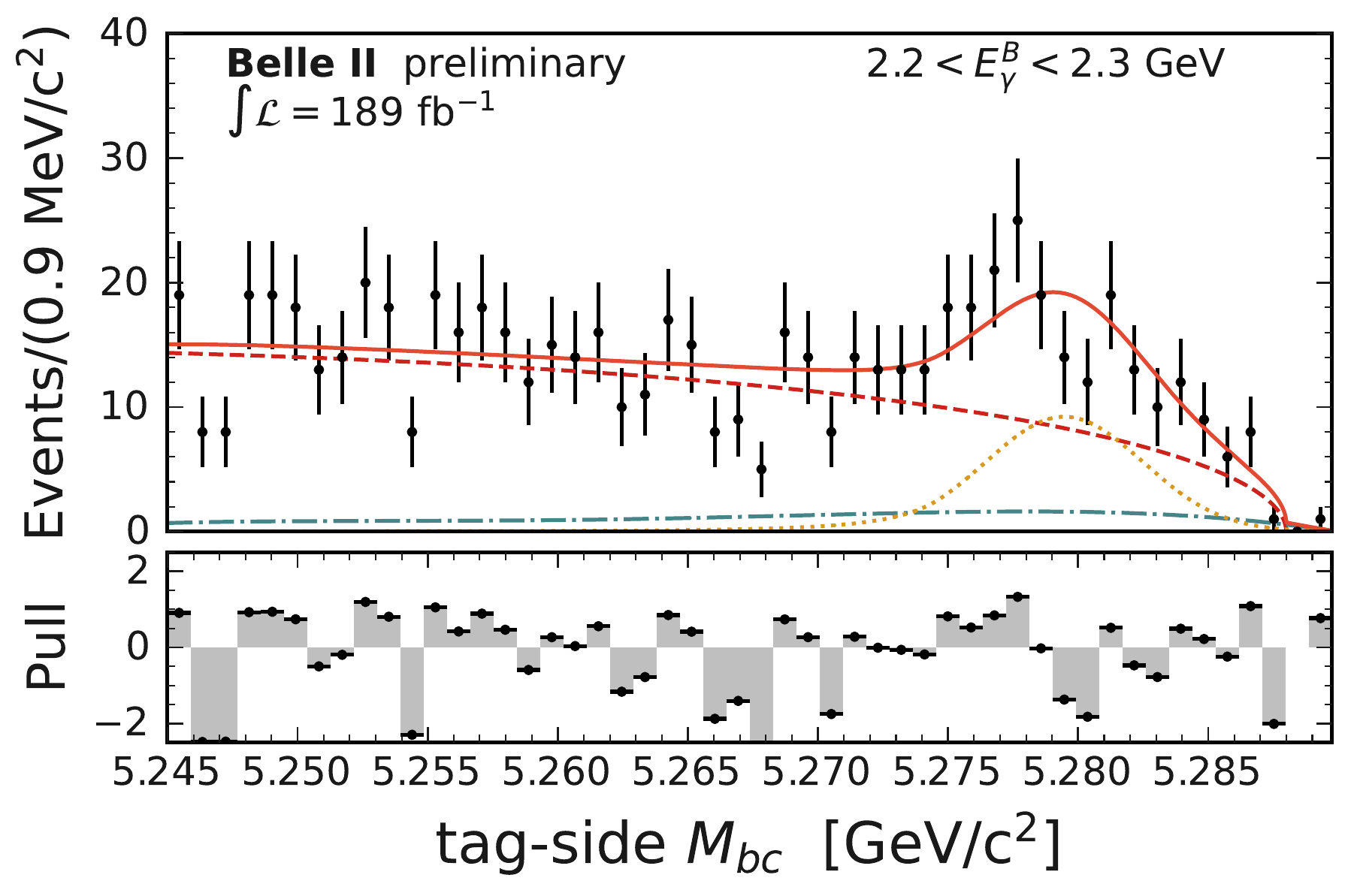}
    \includegraphics[width=0.45\textwidth]{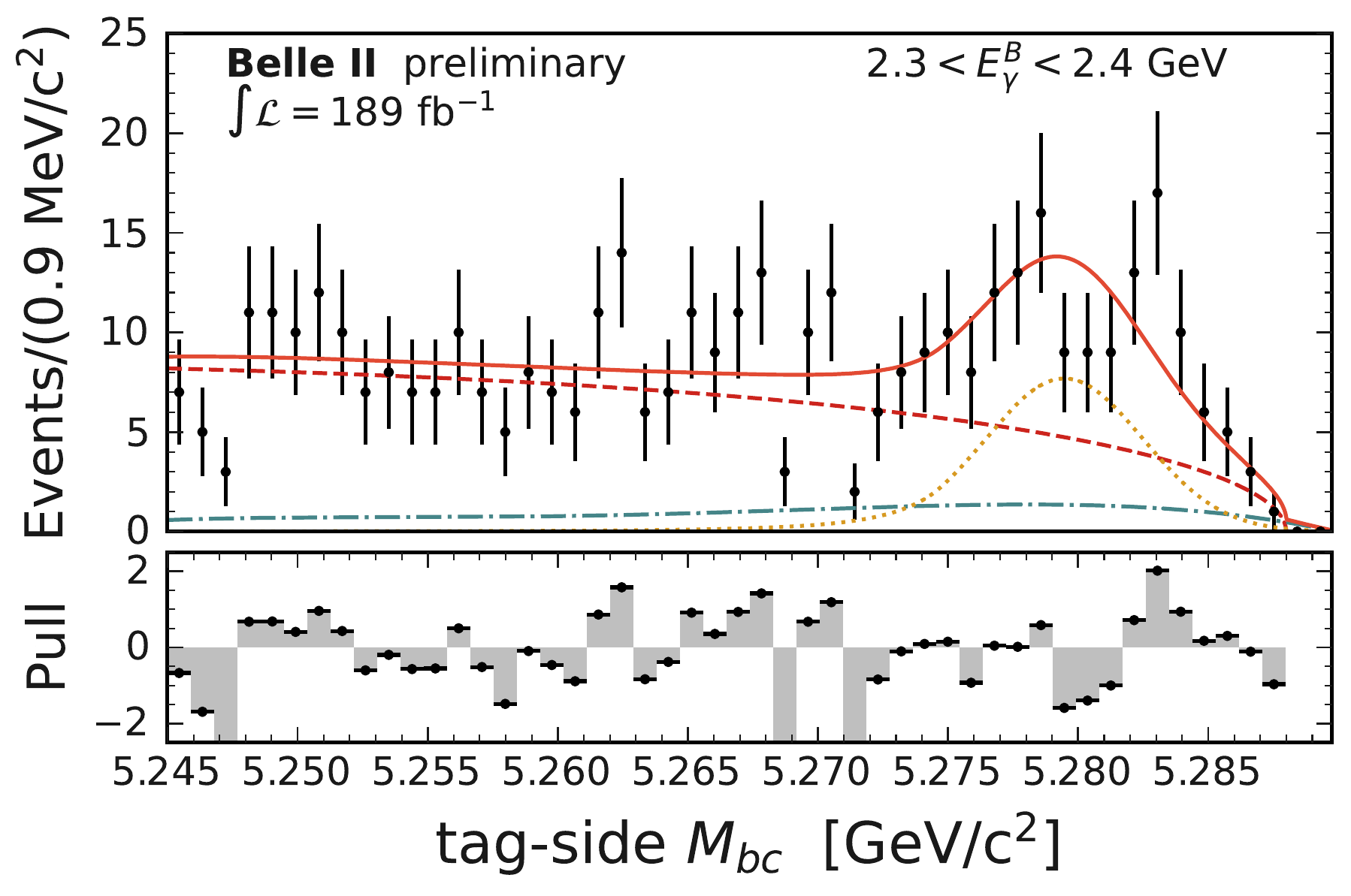}
    \includegraphics[width=0.45\textwidth]{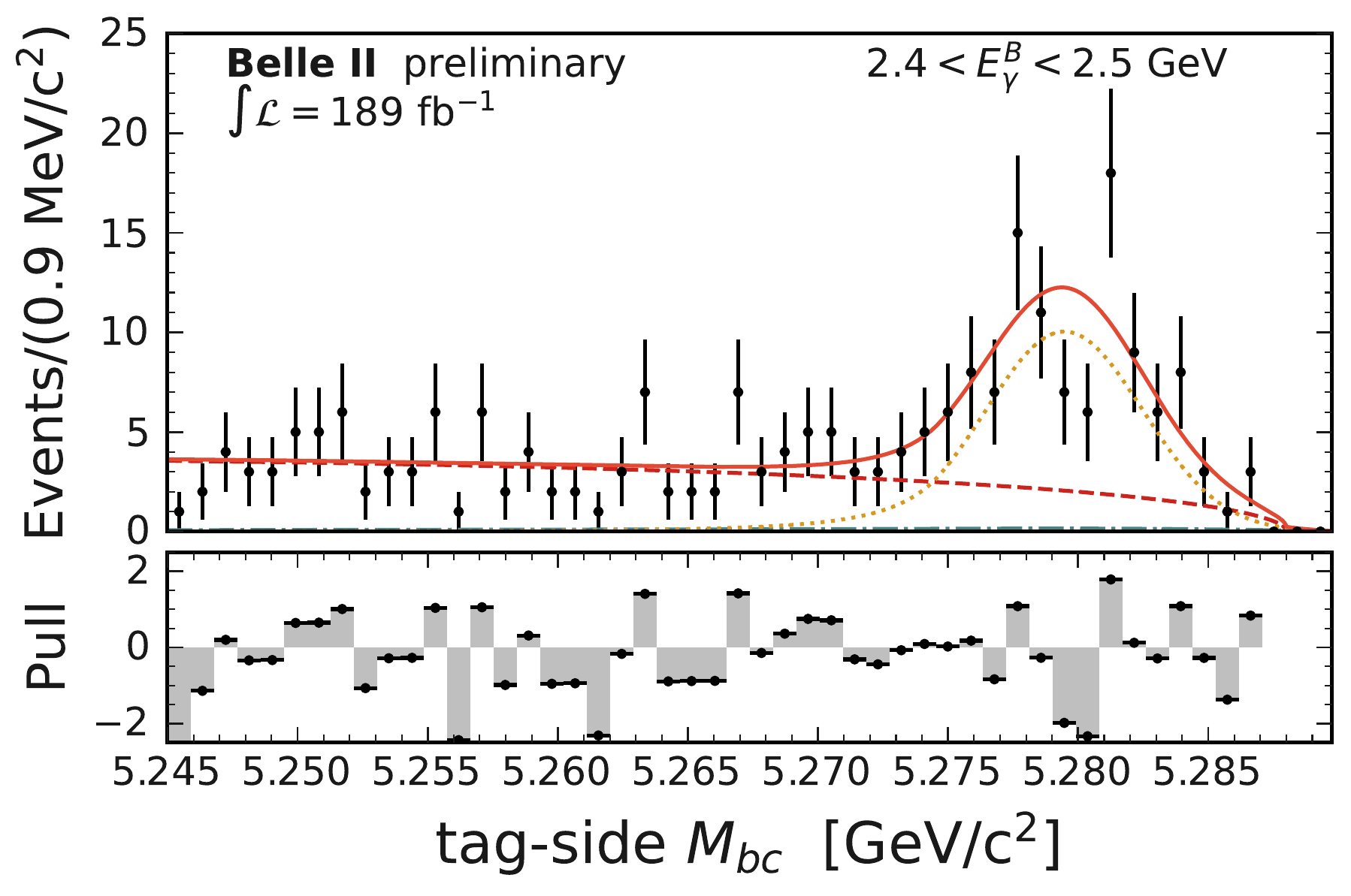}
    \includegraphics[width=0.45\textwidth]{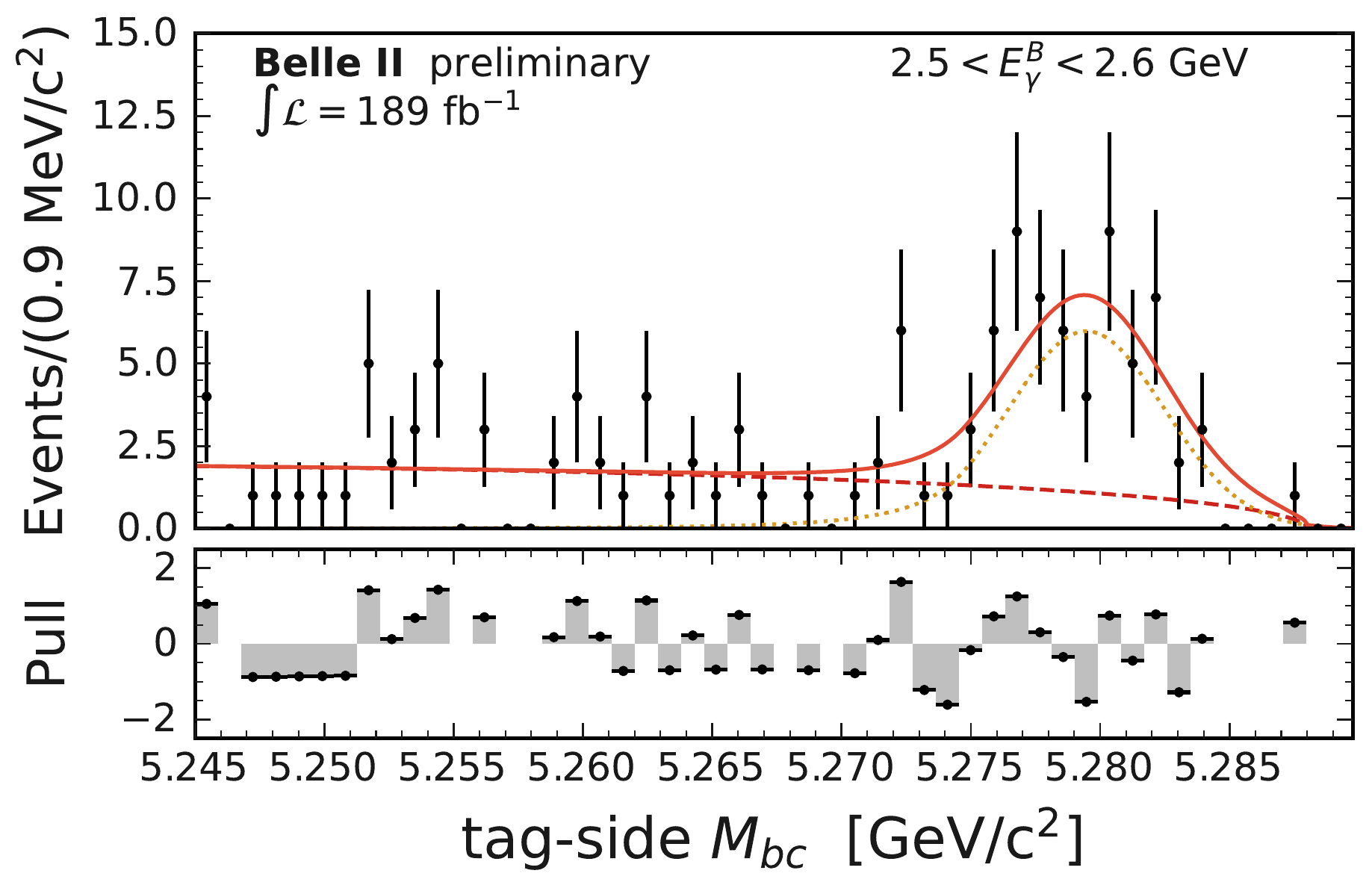}
    \includegraphics[width=0.45\textwidth]{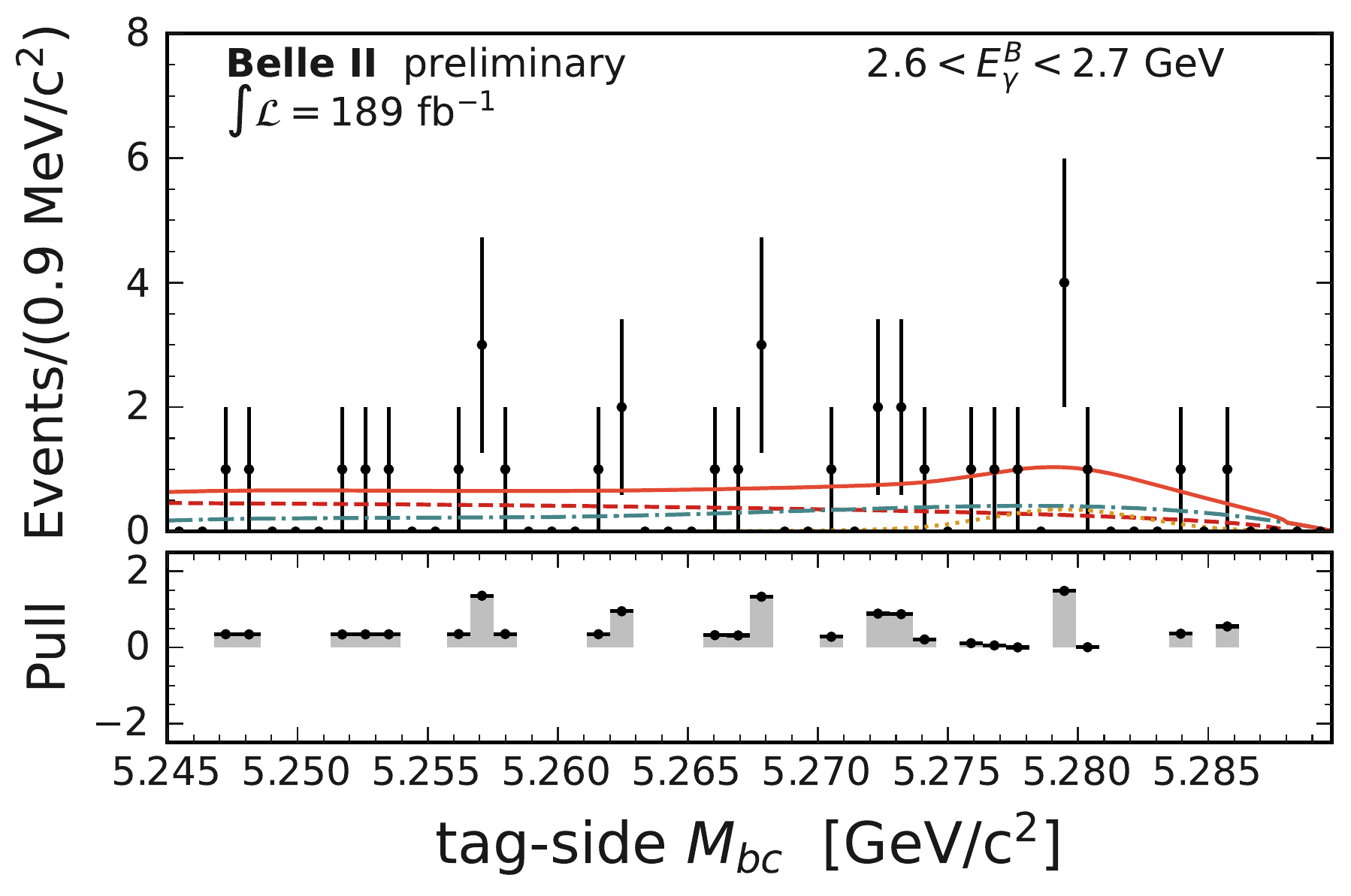}

    \caption{\label{fig:data_fits} Distributions of (black markers with error bars) beam-constrained mass for tag-side $B$ meson candidates restricted to eight \EB bins, with (curves) fit projections overlaid. The orange dotted curve corresponds to the \BB peaking tags. The dashed and dash-dotted curves correspond to the \qqbar and misreconstructed \BB components, modelled by ARGUS and Chebyshev PDFs, respectively. The solid red curve corresponds to the total fit. The lower panels show the difference between fit results and measured values, divided by its statistical uncertainty (pull).}
    \label{fig:mbc_split_fits}
\end{figure}

Likelihood fits to the unbinned \Mbc distributions are performed simultaneously in the \EB bins~\cite{zfit}. A modeling fit is performed on separate components in simulated data to determine the shape parameters and fix them. The fit is then applied to the experimental data. The yields of the three components in each \EB interval, and the ARGUS shape parameters -- which are shared across bins -- are determined by the fit (\Cref{fig:data_fits}). The peaking-$B$ yields in each \EB bin are extracted from the Crystal-Ball normalizations. The peaking-$B$ yield estimator is unbiased and has Gaussian uncertainties as shown by checks on simplified simulated experiments.


\section{Residual \B background subtraction} \label{sec:post_fit}

The resulting peaking $B$ yields include contributions from $\B\ra X_{s+d}\g$ events and other correctly-tagged \BB processes, which are considered background. Due to the high-purity of the tagged sample, background contamination is low at high-\EB, but grows sharply with decreasing \EB. 

\begin{figure}[htbp!]
     \centering
     \includegraphics[width=0.5\textwidth]{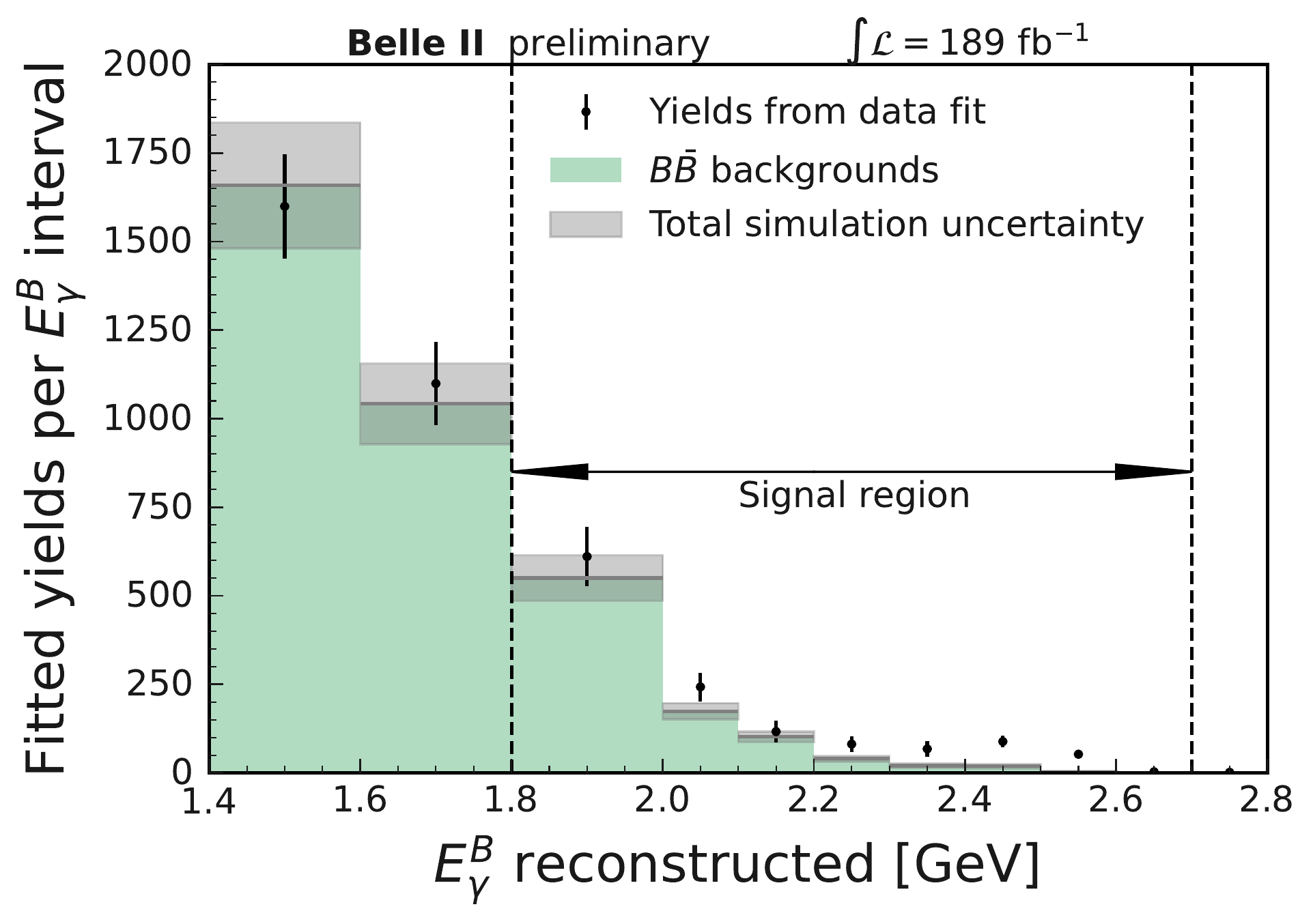}
     \caption{Yield of \BB events as a function of photon energy in the signal \B meson rest frame. The data points correspond to the yields from the fits on the data \Mbc distributions. The histogram shows the luminosity-scaled yields from the background-only simulated sample. The gray bands correspond to systematic uncertainties on the \BB background prediction. The excess of events in data with respect to the \BB background is the $\B\ra X_{s+d}\g$ contribution. }
     \label{fig:background_vs_data}
     
     \vspace{-20pt}
     
 \end{figure}

To remove this background, the PDFs defined in \Cref{sec:fitting_procedure} are used to fit simulated \BB samples in which $\B\ra X_{s+d}\g$ events are removed. This procedure extracts yields of peaking nonsignal events in every \EB bin. The background predictions are scaled for luminosity, and corrected based on FEI calibration factors \cite{Belle-II:2020fst}, \g~detection-efficiency \cite{gamma_eff}, and \piz~efficiency. Branching fractions used in simulation for the background modes are matched to the most recent known values. The \mbox{$1.4<\EB<1.8\gev$} region, where signal purity is low, is used for validating the background subtraction. Event yields observed in data after subtraction are compared with expectations from background-only simulation. An 8.7\% difference is observed, which is assigned as a uniform correction factor for background normalisation across bins. The background expectations from simulation and observed yields in data are shown in \Cref{fig:background_vs_data}.
 
\section{Unfolding}\label{sec:unfolding}

The measured \BtoXsgamma spectrum needs to be corrected (unfolded) for smearing effects. The unfolding uses bin-by-bin multiplicative factors based on the hybrid model. These factors are defined as the ratios between the expected number of events of the generated spectrum and the expected number of corresponding events of the reconstructed spectrum within an \EB interval. The measured \BtoXsgamma yields are multiplied by the unfolding factors (see \Cref{sec:results}). The bulk of the unfolding factors do not exceed 10-20\%, and only the edge bins have 30-60\% corrections.

\section{Uncertainties}\label{sec:uncertainties}

Multiple sources of systematic uncertainty are considered and are grouped as follows: uncertainties due to assumptions in the fit; uncertainties affecting the signal efficiency estimation; data-MC normalisation in the background estimation; and other sources, such as unfolding procedure, branching fraction normalization and the subtraction of $\B\ra X_d \gamma$ component. The statistical uncertainties of the yields extracted from the fit on data are dominant.

\subsection{Uncertainties due to assumptions in the fit}\label{sec:data_uncertainty}

To account for assumptions on the values of model parameters, we repeat the fits by varying the Chebyshev polynomial coefficients by their one-standard-deviation uncertainties, and take the maximum shift in signal yield as the uncertainty. We account for a known data-simulation mismodelling of the \Mbc endpoint due to non-simulated run-dependant variations of the collision energy. The signal yields observed in data using alternative models of background shapes with various \Mbc endpoints are compared. The maximum variation with respect to the central result is taken as uncertainty.

\subsection{Signal efficiency uncertainties}\label{sec:signal_uncertainty}

The signal efficiency is calculated using the simulated hybrid-model signal sample. The values are corrected using FEI simulation-to-data calibration factors, $\mathcal{C}_{\Bz/\Bp}$ \cite{Belle-II:2020fst}, as well as acceptance corrections from the \piz veto and \g efficiency. The corresponding uncertainties related to these factors are propagated as systematic uncertainties. The signal efficiency is validated in the high-purity $\EB \in [2.5, 2.6]~\gev$ region, and the observed difference assigned as an uncertainty. 

\subsection{Background uncertainties}\label{sec:background_uncertainty}

The uncertainties associated with the limited size of the simulated samples used in the \Mbc fits are propagated to the final results. Similarly to the signal efficiency, the background yields extracted from the fits on simulated samples are corrected using FEI calibration, \piz veto efficiency, and \g detection-efficiency correction factors. Uncertainties on the branching fractions of background decay modes are also included. The observed background-normalisation difference (see \Cref{sec:post_fit}) is assigned as a 100\% systematic uncertainty.

\subsection{Other uncertainties}\label{sec:other_uncertainty}

To unfold the measured \EB spectrum, we evaluate the hybrid-model shape uncertainties by taking into account the uncertainty on the ratio of the known branching fractions of $\B\ra\Kstar\g$ to that of  \BtoXsgamma. The \BtoXsgamma model-parameter uncertainties, based on Ref.~\cite{simba}, are also included. The analysis does not distinguish between $X_s$ and $X_d$ final-states. The contribution from the $B\ra X_d\gamma$ component is subtracted assuming the same shape and selection efficiency as \BtoXsgamma. Under this assumption, the $\mathcal{B}(\BtoXsgamma)$ and $\mathcal{B}(\B\ra X_d\g)$ ratio equals $\left|{V_{td}}/{V_{ts}}\right|^2$. The full size of the $B\ra X_d\g$ component is assigned as an uncertainty. The uncertainty on the number of $B$ meson pairs, used as the branching-fraction normalization, is also taken into account. It is estimated by an independent study with a data-driven method in which off-resonance data are used to subtract the non-\BB contribution from the on-resonance data.

\section{Results}\label{sec:results}

The partial branching fractions in the various \EB intervals are calculated as

\begin{equation}\label{eq:branching_fraction}
    \frac{1}{\Gamma_B}\frac{d\Gamma_i}{dE^B_{\g}} = \frac{\mathcal{U}_i\times(N^{\mathrm{DATA}}_i - N^{\mathrm{BKG,~MC}}_i-N_i^{B\rightarrow X_d \gamma})}{\varepsilon_i \times N_B},
\end{equation}

where

\begin{itemize}
    \item $N^{\mathrm{DATA}}_i$ is the peaking-$B$ yield extracted from fitting the data distributions,
    \item $N^{\mathrm{BKG,~MC}}_i$ is the non-$\B\ra X_{s+d}\g$ peaking-$B$ yield expectation extracted from fitting simulated distributions, scaled for luminosity and corrected as discussed in \Cref{sec:post_fit},
    \item $N_i^{B\rightarrow X_d \g}$ is the number of $B\rightarrow X_d \g$ events, equal to $|{V_{td}}/{V_{ts}}|^2 \approx 4.3 \% $ \cite{Workman:2022ynf} of  $N_i^{B\rightarrow X_s \g}$, assuming the same shape and selection efficiency as \BtoXsgamma,
    \item $\varepsilon_i$ is the \BtoXsgamma selection and tagging efficiency, calculated using the simulated hybrid-model sample,
    \item $\mathcal{U}_i$ is the bin-by-bin unfolding factor calculated using the simulated hybrid-model sample,
    \item $N_B\equiv2\times(198 \pm 3)\times10^6$ is the number of \B mesons in the $189~\invfb$ data sample,
    \item $\Gamma_B$ on the left-hand-side of \Cref{eq:branching_fraction} signifies the total decay width of the $B$-meson.

\end{itemize}

\begin{figure}[htbp!]
    \centering
    \includegraphics[width=0.5\textwidth]{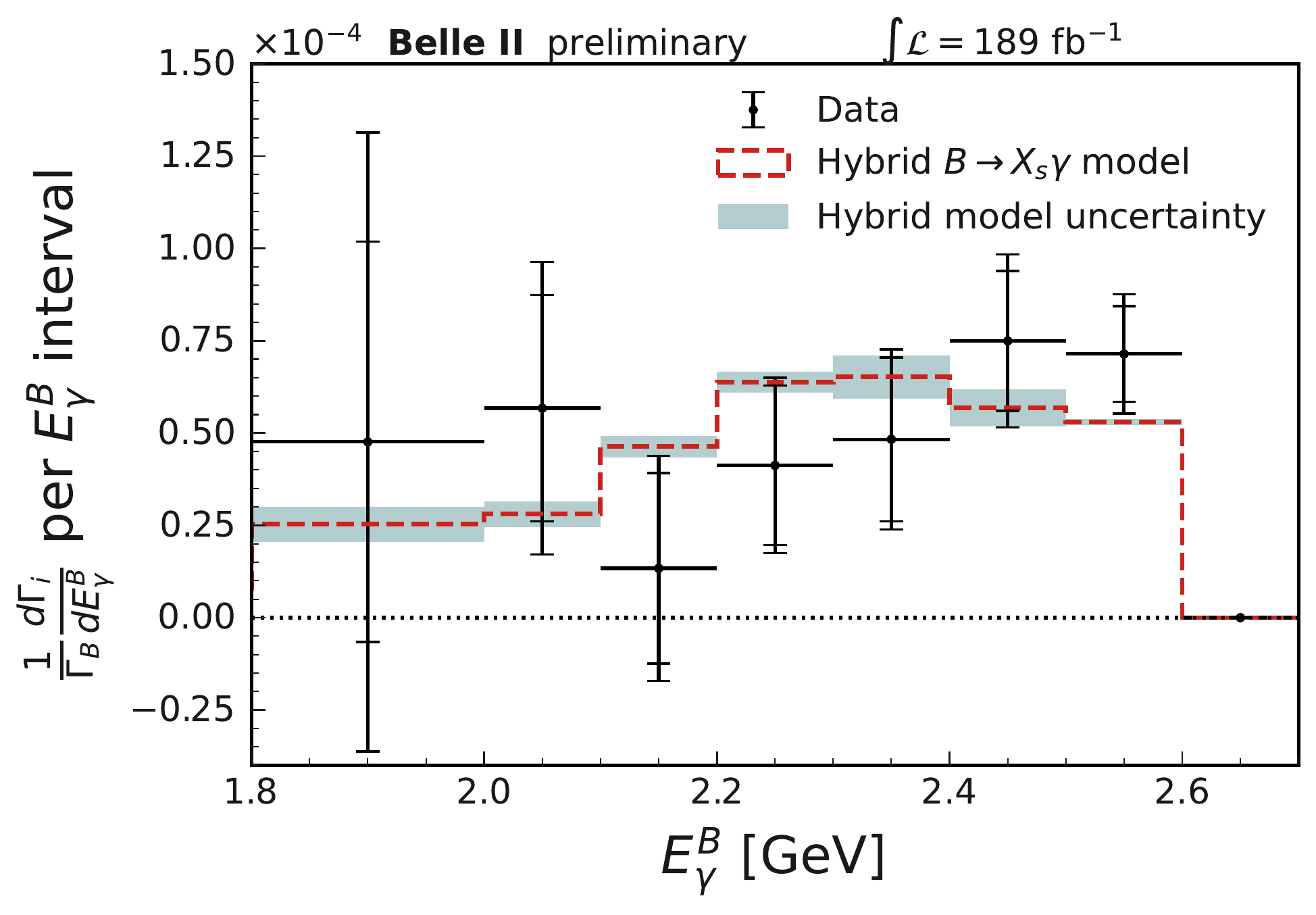}
    \caption{Measured partial branching fractions $ (1/\Gamma_B)(d\Gamma_i/dE^B_{\g})$ as a function of \EB. The outer (inner) uncertainty bar shows the total (statistical) uncertainty. The overlaid model and uncertainty corresponds to the hybrid model.}
    \label{fig:partial_bfs}
\end{figure}

The resulting partial branching fractions are shown in \Cref{fig:partial_bfs}. The various contributions from the major sources of systematic uncertainties as functions of \EB are shown in \Cref{tab:partial_bfs}.

\begin{table}[htbp!]
    \centering
    \caption{Results of the partial branching fraction measurements. The right-hand part of the table shows the main contributions to the systematic uncertainty. Signal efficiency and background modelling uncertainties are correlated (see \Cref{sec:signal_uncertainty,sec:background_uncertainty}).}
    \label{tab:partial_bfs}
    
\resizebox{1\textwidth}{!}{
\begin{tabular}{cccc|cccc}
\toprule
\EB [\gev] &         $\frac{1}{\Gamma_B}\frac{d\Gamma_i}{dE^B_{\g}} (10^{-4})$ &         Statistical &         Systematic & \makecell{Fit \\ procedure} & \makecell{Signal \\ efficiency} & \makecell{Background \\ modelling} &  Other\\
\midrule
$1.8-2.0$  &  0.48 & 0.54 & 0.64 & 0.42 & 0.03 & 0.49 & 0.09 \\
$2.0-2.1$  &  0.57 & 0.31 & 0.25 & 0.17 & 0.06 & 0.17 & 0.07 \\
$2.1-2.2$  &  0.13 & 0.26 & 0.16 & 0.13 & 0.01 & 0.11 & 0.01 \\
$2.2-2.3$  &  0.41 & 0.22 & 0.10 & 0.07 & 0.05 & 0.04 & 0.02 \\
$2.3-2.4$  &  0.48 & 0.22 & 0.10 & 0.06 & 0.06 & 0.02 & 0.05 \\
$2.4-2.5$  &  0.75 & 0.19 & 0.14 & 0.04 & 0.09 & 0.02 & 0.09 \\
$2.5-2.6$  &  0.71 & 0.13 & 0.10 & 0.02 & 0.09 & 0.00 & 0.04 \\
\bottomrule
\end{tabular}
}
\end{table}

The integrated branching ratios for various \EB thresholds are calculated and shown in \Cref{tab:integrated_bfs}. The systematic uncertainties are computed taking the bin-by-bin correlations into account.

\begin{table}[htbp!]

    \caption{Integrated partial branching fractions for three \EB thresholds. The number of observed events before unfolding and efficiency corrections are also given for each threshold.}
    \label{tab:integrated_bfs}

\resizebox{0.9\textwidth}{!}{
\begin{tabular}{ccccc}
\toprule
     \EB threshold [\hspace{-1pt}$\gev$] & & $\mathcal{B}(B\rightarrow X_s \gamma)~[10^{-4}]$ &  & Observed signal yield (tot. unc.) \\
     \midrule
        1.8  & & $3.54 \pm 0.78$ (stat.) $\pm~0.83$ (syst.) & & $343 \pm 122$\\ 
        2.0  & & $3.06 \pm 0.56$ (stat.) $\pm~0.47$ (syst.) & & $285 \pm 68\phantom{0} $\\
        2.1  & & $2.49 \pm 0.46$ (stat.) $\pm~0.35$ (syst.) & & $219 \pm 50\phantom{0} $\\
        \toprule
\end{tabular}

}
\end{table}

\section{Conclusion}

We present a measurement of the photon-energy spectrum in the $B$ meson rest frame from \BtoXsgamma decays using hadronic-tagging of the partner $B$ meson. We also report the inclusive branching ratio $\mathcal{B}(\BtoXsgamma)$ for various thresholds, starting at $\EB>1.8~\gev$. The results are consistent with the Standard Model and world averages \cite{HFLAV:2019otj}. 

\section*{Acknowledgments}

We thank the SuperKEKB group for the excellent operation of the accelerator; the KEK
cryogenics group for the efficient operation of the solenoid; and the KEK computer group for on-site computing support.

\bibliographystyle{JHEP}
\bibliography{references}

\end{document}